\documentstyle[prd,aps,preprint,epsf]{revtex}

\def\NPB#1#2#3{Nucl. Phys. {\bf B#1}, #2 (#3)}

\def\PLB#1#2#3{Phys. Lett. {\bf B#1}, #2 (#3)}

\def\PRD#1#2#3{Phys. Rev. {\bf D#1}, #2 (#3)}

\def\PRL#1#2#3{Phys. Rev. Lett. {\bf#1}, #2 (#3)}
\def\PRT#1#2#3{Phys. Rep. {\bf#1}, #2 (#3)}

\def\JHEP#1#2#3{J. High Energy Phys. {\bf #1}, #2 (#3)}

\def\vev#1{\langle #1 \rangle}
\newcommand{\GEV}{ {\rm GeV} }
\newcommand{\gsim}{ \mathop{}_{\textstyle \sim}^{\textstyle >} }
\newcommand{\lsim}{ \mathop{}_{\textstyle \sim}^{\textstyle <} }

\begin{document}

\tighten
\draft
\title{Vacuum Instability \\
in Anomaly Mediation Models with Massive
Neutrinos}
\author{Masahiro Kawasaki}
\address{Research Center for the Early Universe, University of Tokyo,
   Tokyo, 113-0033, Japan}
\author{T. Watari}
\address{Department of Physics, University of Tokyo, Tokyo 113-0033,
Japan}
\author{T. Yanagida}
\address{Department of Physics, University of Tokyo, Tokyo 113-0033,
Japan \\ and \\
Research Center for the Early Universe, University of
Tokyo, Tokyo, 113-0033, Japan}

\date{\today}

\maketitle

\begin{abstract}

We study the vacuum stability in the anomaly mediated supersymmetry
(SUSY) breaking models with massive neutrinos.  It is shown that,
because of the seesaw-induced mass terms for neutrinos, the true
vacuum has a large negative cosmological constant provided that the
vacuum where we now live has an (almost) vanishing cosmological
constant.  Although the quantum transition into the true vacuum from
our false vacuum is highly suppressed, the thermal transition at high
temperatures may not be neglected because of the thermal excitations.  
However, we find that the thermal
transition is, in fact, negligibly small and hence the anomaly
mediation models are cosmologically safe.  Thus, we conclude that the
reheating temperature $T_R$ could be very high $ (e.g. T_R \gg 10^{10}
\GEV)$ in the anomaly mediation models even with the seesaw-induced mass 
 terms for neutrinos.

\end{abstract}



\section{Introduction}

Recent Superkamiokande experiments on atmospheric
neutrinos~\cite{SuperK} have presented very convincing evidence for
the oscillation of $\nu_{\mu}$ to $\nu_{\tau}$ with a mass difference
$\delta m^2 \simeq 10^{-3} - 10^{-2}$~eV$^2$.  If neutrinos are indeed
massive, the seesaw mechanism~\cite{SeeSaw} is the most natural
framework to account for the smallness of neutrino masses.

We point out, in this paper, that the seesaw-induced mass terms for
neutrinos generate instability of the vacuum in the
supersymmetric(SUSY) standard model, if the anomaly
mediation~\cite{Randall-Sundrum,Giudice-etal} of SUSY breaking
provides the dominant contribution to all the SUSY-standard-model
fields.  The transition of our false vacuum to the true vacuum is
strongly suppressed at zero temperature, since the true minimum is
separated far away from our false vacuum.  However, the thermal
transition seems to be effective at high temperatures because of the
thermal excitations, which may lead
to horrible universes.  We find, contrary to the above thought, that
the thermal transition is also strongly suppressed and hence the
anomaly mediation models are perfectly consistent with the present
observation.  Therefore, the reheating temperature after inflation
would not be constrained from above in anomaly mediation models with the 
seesaw-induced mass terms for neutrinos.  This
our observation makes the anomaly mediation models very attractive,
since most of baryo \& lepto-genesis models need sufficiently high
temperature $T_{R}\gtrsim 10^{10}$~GeV~\cite{Buchmuller}.  Notice that
the anomaly mediation models predict the large gravitino mass $m_{3/2}
\simeq 100$~TeV avoiding the ``gravitino problem'', which is a
serious problem in gravity-mediation models~\cite{Ellis}.

\section{Seesaw Mechanism in Anomaly Mediation Model}

In the anomaly mediation models the SUSY breaking in the hidden sector
is transmitted to the observed sector by super-Weyl anomaly
effects~\cite{Randall-Sundrum,Giudice-etal}. In particular, the
anomaly effects provide the dominant contribution to gaugino masses as
\begin{equation}
     m_{G_{a}} = \frac{b_{a}g_{a}^2}{16\pi^2}\langle \Phi \rangle
|_{\theta^2}
     ~~~~~~(a = 1,2,3),
\end{equation}
where $g_{a}~(a=1,2,3)$ are the gauge coupling constants for
$U(1)_{Y}, SU(2)_{L}$ and $SU(3)_{C}$ gauge groups, and $b_{a}~(a
=1,2,3)$ are $\beta$-function coefficients of the corresponding gauge
coupling constants.  $\Phi$ is the supergravity auxiliary
supermultiplet whose $\theta^2$ component's vacuum-expectation value
(vev) is of order the gravitino mass $m_{3/2}$.  It is very crucial in
the present analysis that the gaugino masses are one-loop suppressed
relative to the gravitino mass and hence the gravitino is extremely
heavy.  For the gluino mass $m_{G_{3}}\simeq 1$~TeV we see the
gravitino mass $m_{3/2} \simeq 100$~TeV. The SUSY-breaking (mass)$^2$
for scalar bosons are induced at the two-loop level.  However, pure
anomaly mediation predicts slepton (mass)$^2$ to be negative,
requiring additional contributions to the slepton masses.  In the
present analysis we employ the simple phenomenological
solution~\cite{Gherghetta} to this problem that merely adds a
universal mass term $m_{0}^2$ of order the
electroweak scale (100~GeV -- 1~TeV)$^2$ to all of the scalar masses,
leaving the gaugino and 
gravitino masses unchanged.  We assume, in this paper, the gravitino
mass $m_{3/2}\simeq 100$~TeV.

Let us now discuss the seesaw-induced mass terms for neutrinos. The
integration of heavy right-handed neutrinos generates the following
non-renormalizable operator in the low-energy superpotential:
\begin{equation}
     W = \frac{1}{M_{R i}} (L_{i}H_{u})^2,
\end{equation}
where $L_{i}~(i=1,2,3)$ and $H_{u}$ are chiral supermultiplets for
lepton and Higgs doublets and $i$ denotes family index.  $M_{R i}$
represent the effective Majorana masses for right-handed neutrinos.
We take, in the present analysis, $M_{R 3} \simeq (0.3 - 1)\times
10^{15}$~GeV, reproducing $m_{\nu_{3}} \simeq (0.3 - 1)\times
10^{-1}$~eV suggested from the Superkamiokande experiments on the
atmospheric neutrino oscillation~\cite{SuperK}.  We suppress the
family indices in the following discussion.

We consider the following $D$-term flat direction:
\begin{equation}
     L = \frac{1}{\sqrt{2}}\left(
            \begin{array}{c}
               \phi \\
	        0
	   \end{array}
	 \right),~~~~
     H_{u} = \frac{1}{\sqrt{2}}\left(
            \begin{array}{c}
                 0 \\
	      \phi
	   \end{array}
	 \right),
\end{equation}
and others $=0$. Then, the superpotential is written as
\begin{equation}
     W = \frac{1}{4M_{R}}\phi^4.
\end{equation}
The scalar potential is given by

\begin{equation}
     \label{eq:potential}
     V = m_{\phi}^2 |\phi |^2
     - \frac{m_{3/2}}{4M_{R}}(\phi^4 + h.c.)
     + \left|\frac{1}{M_{R}}\phi^3\right|^2,
\end{equation}
where $m_{\phi}^2$ is the soft SUSY breaking mass for $\phi$.  Note
that we have assumed our desired vacuum $\vev{\phi} = \phi_F \equiv 0$ to
have an (almost) vanishing cosmological constant.  We easily see that the
desired minimum $\langle \phi \rangle = 0$ is no longer the absolute
minimum in the theory of anomaly mediation with $m_{3/2} \simeq 100 m_{\phi}
$ much larger than $ m_\phi $.  The true vacuum appears
at\footnote{
This (true) vacuum can not be identified with the vacuum we live in,
since the electroweak gauge bosons are too heavy as $m_{W}\simeq
m_{Z} \sim 10^{10}$~GeV there.}
\begin{equation}
     \vev{\phi} \simeq \phi_T \equiv
     \sqrt{ \frac{m_{3/2}M_{R}}{3}}\sim 10^{10}~\GEV ,
\end{equation}
and it has a negative cosmological constant given by

\begin{equation}
     V_{\rm true} = -\frac{m_{3/2}^3M_{R}}{54}.
\end{equation}
Since the potential energy $V( \phi = 0) \simeq 0$ of
the present universe is much higher than $V_{\rm true}$, we live in
the false vacuum now\footnote{
It may be possible that the $H_u = L $ flat direction is lifted up 
in some extended models by introducing new particles at intermediate
energy scales. In those cases, the true vacuum with a large negative
cosmological constant may not appear.}.

This false vacuum is in principle unstable against tunneling into the
true vacuum.  If the anomaly mediated SUSY breaking models really
describe our world, the tunneling rate should be very small in order
that the present false vacuum survives at least longer than the
age of the present universe.  Furthermore, we must have had some
certain cosmological history that has safely led us to this false vacuum.

\section{Vacuum Transition Rate}

\begin{figure}[t!]
\centering
\hspace*{-7mm}
\leavevmode\epsfysize=14cm \epsfbox{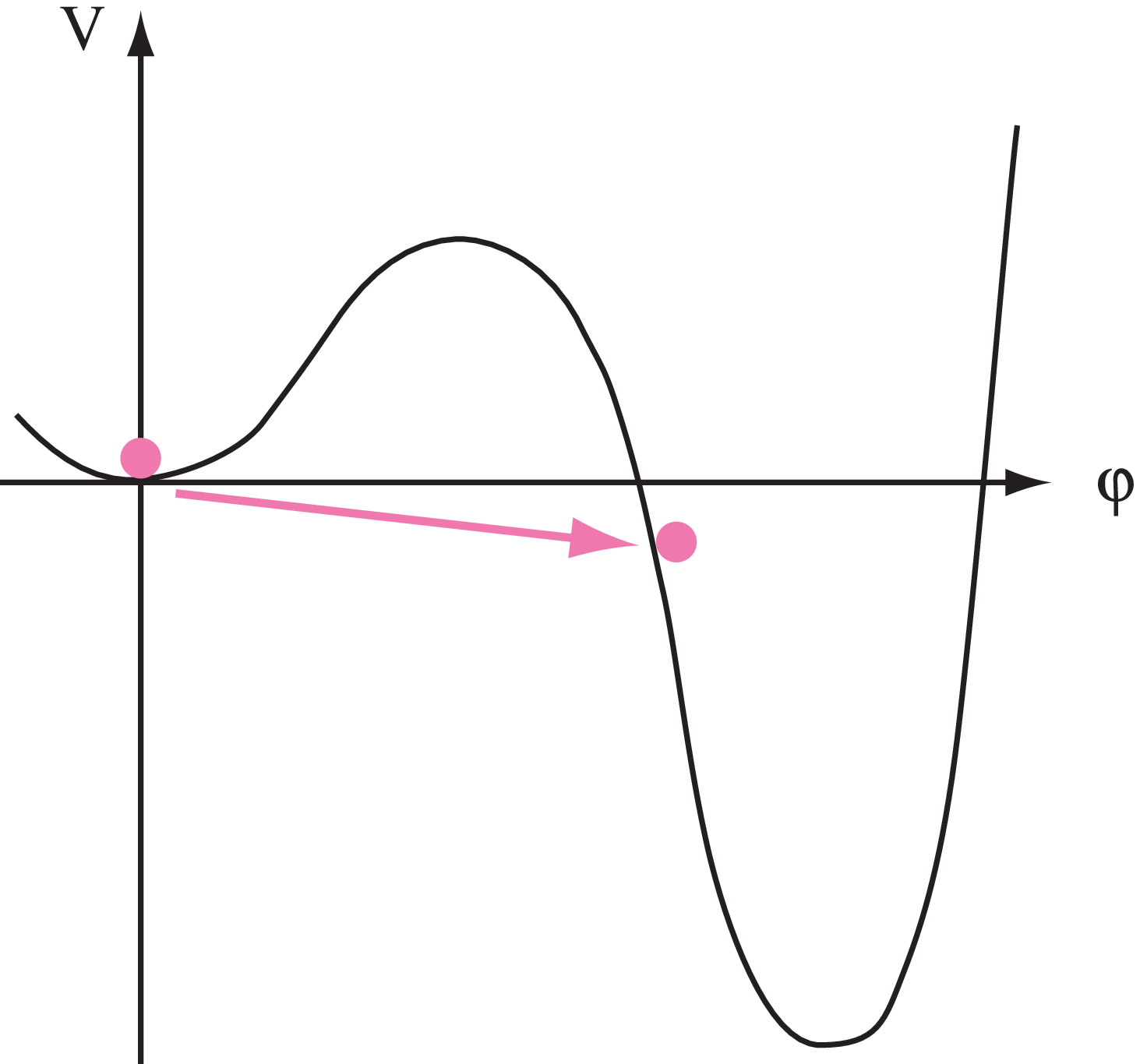}\\[2mm]
\caption{\label{fig:potential}
Potential for $\varphi$.}
\end{figure}

Let us estimate the tunneling rate of our false vacuum.  At zero
temperature we must estimate four dimensional Euclidean action $S_{4}$
evaluated with the bounce solution for the potential
(\ref{eq:potential})~\cite{Coleman}.  Since the potential at the true
minimum is deep in comparison with the height of the potential barrier
[$ i.e. ~ V_{\rm barrier} \sim (m_{\phi}/m_{3/2})^4|V_{\rm true}| \ll
|V_{\rm true}|$], we can
neglect the $\phi^6$-term in the potential for $|\phi| \lsim \phi_{\rm
barrier} \sim \sqrt{(m_\phi/m_{3/2})} \phi_T$ ({\it i.e.} the thick wall
approximation) 
[see Fig.~\ref{fig:potential}].  Then $S_{4}$ is given by

\begin{eqnarray}
     S_{4}  & = & \int dx^4 \left(
     \frac{1}{2} \left(\frac{d\varphi}{dt}\right)^2 +
     \frac{1}{2}(\nabla \varphi)^2 + V(\varphi)\right),\\
     V(\varphi) & \simeq & \frac{1}{2}m_{\phi}^2 \varphi^2
     -\frac{1}{8} \frac{m_{3/2}}{M_{R}}\varphi^4,\label{eq:def-lambda}
\end{eqnarray}
where $\varphi =  Re\phi / \sqrt{2}$ and $\varphi$ satisfies
\begin{equation}
     d^2\varphi/dt^2 + \nabla^2 \varphi = dV/d\varphi.
\end{equation}

By redefining $\varphi = m_{\phi}(4 M_{R}/m_{3/2})^{1/2}\psi,
x_{\mu} = \xi_{\mu}/m_{\phi}$  with $\psi$ and $\xi$ being
dimensionless, we can easily see that $S_{4}$ is rewritten as

\begin{eqnarray}
     S_{4}  & = & \frac{4M_{R}}{m_{3/2}} S_{4}(\psi),\\
     S_{4}(\psi) & \simeq & \int d^4\xi \left(
     \frac{1}{2} (d\psi/d\xi^0)^2 + \frac{1}{2}(\nabla_{\xi}\psi)^2
     + \frac{1}{2}\psi^2 - \frac{1}{2}\psi^4\right),
\end{eqnarray}
where $S_{4}(\psi)$ $\sim 10$~\cite{Linde}. Then we obtain the
tunneling rate $\Gamma_{4}$ as

\begin{equation}
     \Gamma_{4} \sim m_{\phi}^4 S_{4}^2 \exp(-S_{4})
     \sim m_{\phi}^4 \frac{1600 M_{R}^2}{m_{3/2}^2}
     \exp\left(-\frac{40 M_{R}}{m_{3/2}}\right).
\end{equation}
Since $M_{R}/m_{3/2} \sim 10^{10}$, the tunneling rate $\Gamma_{4}$ is
negligibly small and hence the vacuum transition into the true vacuum
is sufficiently suppressed at zero temperature.

We can see that this stability against the tunneling comes from the
smallness of the coefficient $\lambda$ of the quartic term ($\lambda \equiv
1/8 (m_{3/2}/M_R) \simeq 10^{-11} \ll 1$) in the
potential (\ref{eq:def-lambda}). Indeed the exponent is roughly
estimated as
\begin{equation}
  S_4 \sim ({\rm 4D ~bubble ~volume}) \times ({\rm Lagrangian})
      \sim \frac{1}{m_\phi^4} V_{\rm barrier}
\end{equation}
where
\begin{equation}
V_{\rm barrier} \sim \frac{1}{\lambda} m_\phi^4 \gg m_\phi^4,
\end{equation}
and hence
\begin{equation}
  S_4 \sim \frac{1}{\lambda}( = \frac{M_R}{m_{3/2}}) \gg 1.
\end{equation}
The small coefficient has led to the large expectation value $ \phi_T
\simeq (1/\sqrt{\lambda}) m_\phi $ and the large potential barrier
compared with the mass $m_\phi$.


We have found that once the $\vev{\phi}=\phi_F \equiv 0$ vacuum (which
we call $\phi_F$-vacuum hereafter) was chosen and the temperature has
dropped down almost to zero, the quantum tunneling to the $\vev{\phi}=\phi_T
$ vacuum (which we call $\phi_T$-vacuum) is
sufficiently suppressed. Now let us examine  whether or
not this $\phi_F$-vacuum is chosen naturally in the cosmological history.

What can be plausibly taken to be the initial condition of the
cosmological history?  We have many evidences which suggest that there
exists an inflationary epoch before big bang nucleosynthesis (BBN).  This
inflationary epoch is followed by reheating process.
The energy density of the
universe which was initially carried by the inflaton potential energy is
gradually but completely converted into thermal plasma energy through inflaton
decay.
 
During this reheating process, the temperature of the thermal plasma
changes as a function of time as\cite{KolbTurner}
\begin{equation}
   T^4 = 1.2 H \Gamma_{\rm{inf}} M_G^2, \qquad \qquad H =
\frac{2}{3t},
\end{equation}
where $H$ is Hubble parameter during the reheating process,
$\Gamma_{\rm inf}$ decay rate of the inflaton and $M_{G}\simeq
2.4\times 10^{18}$~GeV the reduced Planck mass.  The universe takes
the maximum temperature $T_{m}$ soon after inflation and this
temperature is much higher than that at the end of the reheating
process (the reheating temperature $T_R \sim (\Gamma_{\rm{inf}}^2
M_G^2)^{1/4}$).  In this thermal plasma with high temperature, the
$\phi$ field feels finite temperature effective
potential which is drastically different from the zero temperature
potential (\ref{eq:potential}).


Now let us see this thermal effective potential in detail.
   Particles which have interactions with the $\phi$ field
give contributions to the thermal potential of the $\phi$ field. These
contributions exist as long as those particles are in thermal
equilibrium and therefore as long as masses of those particles are less
than the temperature. The $\phi$ field expectation value $\langle \phi
\rangle$ gives masses to those particles through the superpotential,
\begin{eqnarray}
     W & = & Y_{i}Q_{i}\bar{U}
     \frac{\phi}{\sqrt{2}}
     + Y_{\rm eff}\bar{E}_{\rm eff}H_{d}
     \frac{\phi}{\sqrt{2}}, \\
     & & Y_{\rm eff} = \sqrt{\sum_{f}|Y_{f}U_{f3}|^2},\\
     & & \bar{E}_{\rm eff} =
     \frac{1}{\sqrt{\sum_{f}|Y_{f}U_{f3}|^2}}
     \sum_{f}Y_{f}U_{f3}\bar{E}_{f},
\end{eqnarray}
where $Y_{i}$ denote up-type Yukawa couplings, $Y_{f}$ denote
charged-lepton Yukawa couplings, $U_{f3}$ denote lepton-flavor mixing
matrix elements, and $H_{d}$ is Higgs supermultiplet couple to
down-type quarks.  $Y_{i}$ and $Y_{f}$ are given by
\begin{eqnarray}
     Y_{i} & = & m_{i}/(174\GEV\sin\beta )
     ~~~ (i= {\rm top}, {\rm charm}, {\rm up}), \\
     Y_{f} & = & m_{f}/(174\GEV \cos\beta )
     ~~~ (f= \tau, \mu, e),
\end{eqnarray}
where $\beta = \arctan(\langle H_{u} \rangle /\langle H_{d} \rangle)$.
Masses of $ Q_i $ and $\bar{U}_i$ supermultiplet particles are $Y_i
\phi/\sqrt{2}$ and those of $\bar{E}_{\rm eff}$ and $H_d$ particles
are $Y_{\rm eff}\phi/\sqrt{2} $.  Hence the range of the field value
in which the thermal potentials arise from those particle loops are
roughly limited to $Y |\varphi| \lesssim T $.
The contribution to the thermal potential ($=$ free energy) from each
chiral multiplet is
given by $-T^4 \pi^2/24 + Y^2 \varphi^2 T^2 /8 $.

We now assume, for example, $T_m > 10^{10} \GEV$. In this case, while
the temperature is higher than $10^{10} \GEV$, the top quark multiplet
does not decouple from the thermal equilibrium even for
$\varphi \sim \phi_T \simeq \sqrt{M_R m_{3/2}} \sim 10^{10} \GEV$
(since $m_{\rm top} \simeq Y_t \phi_T < T $).\footnote{
Here and from now on, we assume that $Y_t \sim 1, Y_{c,\tau} \sim
10^{-2}$ and $Y_u \sim 10^{-4}$.}
Then for all $\varphi \lesssim \phi_T$, the thermal mass term $(1/2)
m_{\rm eff}^2 \varphi^2$ arising from top (s)quark multiplet, 
where $m_{\rm eff} \sim Y_t T \gtrsim
10^{10}\GEV \gg m_{3/2}$, is dominant over the negative quartic term which
originally
exists.  There is no local minimum anywhere except for $\varphi =\phi_F=0 $
and the universe sits around the origin $\varphi = 0$.

When the temperature drops below $10^{10} \GEV$, the top (s)quark multiplet
decouples from the equilibrium for $\varphi \sim {\cal O}(\phi_T)$
and this multiplet contributes to the thermal potential only for
$\varphi \ll \phi_T$.  While the temperature is larger than $10^8
\GEV$ (and less than $10^{10} \GEV$), the effective mass term for $
\varphi \sim {\cal O}(\phi_T)$ comes from the thermal contribution
of lepton, Higgs and charm (s)quark multiplets .  This thermal mass term
is also large enough to dominate the negative quartic term
(since $ m_{\rm eff} \sim Y_{c,\tau} T \gtrsim 10^6 \GEV \gg m_{3/2}$) and
hence the $\varphi = 0$ minimum is still the only and absolute
minimum.\footnote{
We notice here that we do not need to assume $T_m > 10^{10}
\GEV$. Assumption of $T_m > 10^8 \GEV$ is sufficient.}

\begin{figure}[t!]
\centering
\hspace*{-7mm}
\leavevmode\epsfysize=14cm \epsfbox{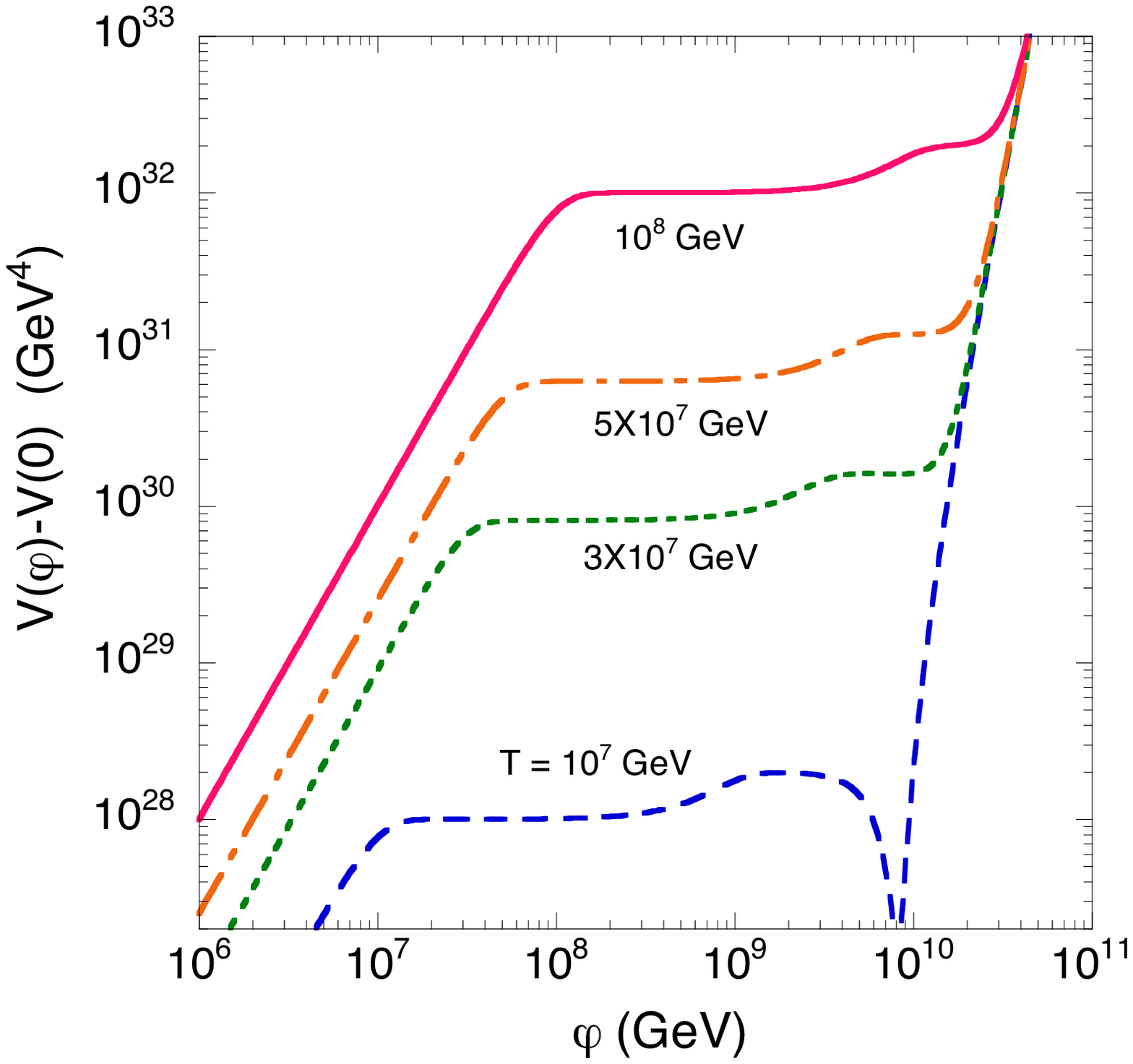}\\[2mm]
\caption{\label{fig:pot-FT}
Finite temperature potential for $\varphi$. The  solid, dash-dotted,  short
dashed and long dashed lines represent the potentials for temperature
$10^8$,$5\times 10^7$,$3\times 10^7$ and $10^7$~GeV, respectively.}
\end{figure}

After the temperature gets down below $10^8 \GEV$, lepton, Higgs and
charm (s)quark supermultiplets no longer give thermal potential for $\varphi
\sim {\cal O}(\phi_T) \sim {\cal O}(10^{10} \GEV)$, since the
temperature is not enough to thermalize those particles with masses
$m_{c,H_d,\bar{E}_{\rm eff}} \sim Y_{c,\tau} \phi_T \sim 10^8
\GEV$.
The thermal potential by now comes only from up (s)quark
supermultiplet,\footnote{
Electroweak gauge bosons have already decoupled from
the thermal equilibrium at this stage.}
  which is so
tiny (since $m_{\rm eff} \simeq \sqrt{(Y_u T)^2+m_\phi^2} \simeq m_\phi \ll
m_{3/2}$) that the effective potential of the field $\phi$ begins to show a
dip around the would-be true minimum $\varphi \simeq \phi_T \simeq 10^{10}
\GEV$(Fig.\ref{fig:pot-FT}).  As the temperature falls further, this dip gets
larger.  The local minimum newly appeared there will become the true
minimum before the temperature drops down to
$10^7\GEV$(Fig.\ref{fig:pot-FT}).

This change in the shape of effective thermal potential is exactly the
same as that in the standard first order phase transition.
Naive guess will tell us that the phase transition to the $\phi_T$-vacuum
occurred in the history of the universe as usual.  We study in the
following whether the phase transition really occurs or not.


First order phase transitions are known to take place through following
  two mechanisms.  One mechanism is through equilibrium between
$\phi_F$- and $\phi_T$-vacua~\cite{Gleiser}, and the other is the conventional
bubble nucleation process~\cite{Linde}.

At first we study 
the phase transition through the
thermal equilibrium between $\phi_F$-vacuum($\vev{\phi}=0$)
and $\phi_T$-vacuum ($ \vev{\phi}\simeq \phi_T \sim 10^{10}\GEV$).
Key idea of this mechanism is as follows.  Once the
dip in the potential is formed then domains of $\phi_F$-vacua
and $\phi_T$-vacua appear.  The size of each domain is roughly the
correlation length ($\sim $ curvature inverse) of the scalar field.
These two types of domains are in the thermal equilibrium via thermal
transition of vacua beyond the potential barrier.  The transition rate
from $\phi_F$-vacuum to $\phi_T$-vacuum and vice versa are given
by~\cite{Gleiser}
\begin{eqnarray}
    \Gamma_{F \rightarrow T} \sim
     m_F^4 e^{- \frac{F_{F \rightarrow T}}{T}} \\
    \Gamma_{F \leftarrow T} \sim
     m_T^4 e^{- \frac{F_{F \leftarrow T}}{T}}
\end{eqnarray}
where $ m_{F,T}$ are curvature of the effective potential at each
vacua, $F_{F \rightarrow T,F \leftarrow T}$ are free energy barrier of
transition in each direction.  The ratio of number of $\phi_F$- and $\phi_T$-
vacua is determined by detailed balance condition if the system is in
the equilibrium:
\begin{eqnarray}
    N_F \Gamma_{F \rightarrow T} & = & \Gamma_{F \leftarrow T} N_T \\
    \frac{N_T}{N_F} & = & \left( \frac{m_F}{m_T}\right)^4
    \exp\left(-\frac{F_{F \rightarrow T}-F_{F \leftarrow T}}{T}\right).
\end{eqnarray}
If the system keeps equilibrium until the $\phi_T$-vacuum really
becomes true vacuum and until the domain size becomes larger than the
critical radius,\footnote{
Characterization of the ``critical radius'' is as follows:
Bubbles with radius larger than the ``critical radius'' expand and
others not.}
then the true ($\phi_T$-)vacuum domains begin to expand and percolate, and the
phase transition is completed.

Now let us examine if the equilibrium is maintained for the case of
our interest.  We are going to make an estimate of the transition rate
from the $\phi_F$- to the $\phi_T$-vacuum.  When the dip in the potential was
formed($T \sim 10^8 \GEV$), the barrier height of the free energy
density in this transition is roughly given by
\begin{equation}
    f_{F \rightarrow T} \sim Y_{c,\tau}^2 T^2 \phi_T^2.
\end{equation}
Since the typical volume of each domain is roughly $ m_{\rm
eff}^{-3}\simeq (1/(Y_{c,\tau} T)^3)$, the barrier of the free energy
is written as
\begin{equation}
    F_{F \rightarrow T} =
    \frac{Y_{c,\tau}^2 T^2 \phi_T^2}
    {(Y_{c,\tau} T)^3} .
\end{equation}
Then, we obtain the transition rate as
\begin{equation}
     \Gamma_{F \rightarrow T} \sim (Y_{c,\tau} T)^4
     \exp\left(
     -\frac{\phi_T^2}{Y_{c,\tau} T^2}
     \right) \sim 10^{-8}T^4 \exp(-{\rm factor}\times 10^5).
\end{equation}
This transition rate is extremely small compared with the expansion
rate of the universe $H^4 \gtrsim T^8 / M_G^4 \sim T^4 10^{-40}$.
Therefore, it is concluded that the transition between the
$\phi_F$-vacuum and the $\phi_T$-vacuum is  already
frozen out, and hence no $\phi_T$-vacuum domain is created.
The phase transition through the equilibrium between $\phi_F$- and
$\phi_T$-vacuum does not take place in the case of our interest.


Even if the thermal equilibrium of two vacua are not formed, the
false($\phi_F$-)vacuum can decay into true ($\phi_T$-)one,
after the potential energy of the $\phi_T$-vacuum becomes lower than
that of the $\phi_F$-one, through the thermal bubble nucleation.
We estimate the thermal transition rate below.  In the case of
finite temperature, the transition rate is determined by
three dimensional Euclidean action $S_{3}$\cite{Linde} which is given by
\begin{equation}
     S_{3}   =  \int d^3 x\left(
     \frac{1}{2}(\nabla \varphi)^2 + V(\varphi)_{\rm thermal}\right),
\end{equation}
where $V(\varphi)_{\rm thermal}$ is the finite temperature  potential and
$\varphi$ satisfies
\begin{equation}
     \nabla^2 \varphi = dV_{\rm thermal}/d\varphi.
\end{equation}
 
In the same way as $S_{4}$, $S_{3}$ is calculated with use of the
dimensionless quantities [$\varphi = m_{\rm eff}(4 M_{R}/m_{3/2})^{1/2}\psi,
x_{\mu} = \xi_{\mu}/m_{\rm eff}$] as
  \begin{eqnarray}
     \label{eq:S3}
     S_{3}  & = & \frac{4m_{\rm eff}M_{R}}{m_{3/2}} S_{3}(\psi),\\
     S_{3}(\psi) & \simeq & \int d^3 \xi \left(
     \frac{1}{2} (\nabla_{\xi}\psi)^2 +
     \frac{1}{2}\psi^2 - \frac{1}{2}\psi^4\right),
\end{eqnarray}
where $S_{3}(\psi)$ $\simeq 9.5$~\cite{Linde}.  Then the transition
rate with temperature $T$ is estimated as
\begin{eqnarray}
     \Gamma_{3} & \simeq & T^4
     \left(\frac{S_{3}}{2\pi T}\right)^{3/2}
     \exp(-S_{3}/T) \nonumber \\
     & \simeq &
     T^4 \left(\frac{ 19 m_{\rm eff}M_{R}}{2\pi T m_{3/2}}
     \right)^{3/2}
     \exp\left(-\frac{19 m_{\rm eff}M_{R}}{T m_{3/2}}\right).
\end{eqnarray}
As mentioned before, the $\phi_T$-vacuum becomes really true vacuum
after the temperature cools down to $\sim 10^{7}$~GeV.
At that time, $m_{\rm eff}$ is already $\simeq m_{\phi}$.
%
Therefore, the exponent is bounded from
below as $10^7$ for $T \lesssim 10^7$~GeV.

Now let us calculate the
fraction $P(t_0)$ in the present universe which remains in the $\phi_F$-vacuum state. 
Assuming that the $\phi_T$-vacuum bubble produced in the transition
expands at the light velocity, we obtain
$P(t_0)$~\cite{Guth-Weinberg,Linde},
\begin{equation}
    \label{eq:fraction}
    P(t_0) = \exp \left[-\int_{t_{i}}^{t_0} dt_{1} \Gamma_{3}(t_1)
    a(t_{1})^3 \left\{\frac{4\pi}{3}
    \left(\int_{t_{1}}^{t_0}\frac{dt_{2}}{a(t_{2})}\right)^3
    \right\}\right],
\end{equation}
where $t_{i}$ is the initial time.  
We assume, for simplicity, 
that the critical temperature below which the $\phi_T$-bubbles are
formed is lower than the reheating temperature,
or in other words, the
thermal bubble nucleation occurs in the radiation dominated 
universe\footnote{
This is mere a technical assumption and is not important. 
}. 
Then, 
\begin{equation}
 a(t_{1})^3 \left\{\frac{4\pi}{3}
    \left(\int_{t_{1}}^{t_0}\frac{dt_{2}}{a(t_{2})}\right)^3
    \right\} \sim \left(\frac{T_0}{T_1}\right)^3 \frac{1}{H_0^3},
\end{equation} 
where $T_0$ and $H_0$ are the temperature and Hubble parameter in the
present universe, respectively. 
Since the transition rate $\Gamma_3$ decreases rapidly as
the temperature falls, the time integration in the eq.(\ref{eq:fraction})
can be replaced by
\begin{equation}
 dt_1 \sim d\left( \frac{M_G}{T^2}\right) \sim d\left(\frac{1}{T}\right) \frac{M_G}{T} \sim \frac{1}{S_3}\frac{M_G}{T_i},
\end{equation}
where the $T_i$ is the initial temperature $10^7 \GEV$.
The total expression of the eq.(\ref{eq:fraction}) is
\begin{eqnarray}
 P(t_0) & \sim & \exp \left[-\left(\frac{T_0}{H_0}\right)^3 \frac{M_G}{S_3} 
                         \left(\frac{S_3}{2 \pi T_i}\right)^{\frac{3}{2}} 
                         \exp \left( - \frac{S_3}{T_i} \right)
		   \right]  \nonumber \\
        & \sim & \exp \left[ - 10^{87} \frac{M_G}{S_3} 
		           \left(\frac{S_3}{2 \pi T_i}\right)^{\frac{3}{2}} 
                           \exp \left( - \frac{S_3}{T_i} \right)
		  \right].
\end{eqnarray}
$P(t_0)$ is almost 1 if the exponent $ S_3/T_i $ is larger than
 200.  Therefore the condition that the most of our universe is in
 the $\phi_F$-vacuum today ({\it i.e.} $1-P(t_0) \ll 1$) 
requires $S_3 / T_i \gsim 200 $, which is satisfied in the present 
case ($S_3 / T_i \sim 10^7$). 



\section{Conclusion}

We have studied the vacuum stability in the anomaly mediated SUSY
breaking models with massive neutrinos.  If the small masses of
neutrinos are generated by the seesaw mechanism, the seesaw-induced
mass terms make the present vacuum ($\phi_F$-vacuum) with an (almost)
vanishing cosmological constant unstable and the true vacuum
($\phi_T$-vacuum) has a disastrously large negative cosmological constant.

Although our false vacuum has quite high energy density compared with
the true vacuum, the quantum transition into the true vacuum is highly
suppressed.  High temperature(more than $10^8 \GEV$) thermal plasma
that is created after inflation gives effective
potential of  the  field $\phi$ with a unique vacuum at the origin 
$\vev{\phi}=0$($\phi_F$-vacuum).  As the temperature decreases
the effective potential shows a new local minimum($\phi_T$-vacuum)
which turns into the true minimum at  $T \lesssim 10^7 \GEV$.
The phase transition to the $\phi_T$-vacuum is also
highly suppressed both through the thermal equilibrium domain
formation of both vacua and through thermal tunneling decay process.
Thus, we can live on a supercooled false vacuum state.

Since the thermal transition from the $\phi_F$-vacuum to the
$\phi_T$-vacuum is negligible, it
does not give any constraint on the reheating temperature after
inflation.  Furthermore, the gravitino mass is so heavy ($\simeq
100$~TeV) in anomaly mediation models that gravitinos decay much
earlier than the BBN and hence the model avoids the ``gravitino
problem'', which is a serious problem in gravity-mediation models.
Thus, the reheating temperature is not constrained from above, which
is very favored by many baryogenesis scenarios.

We have discussed only on the heaviest neutrino direction.  For
lighter neutrino directions the analyses are the same.  We easily see
that the potential barrier between the true minimum and our false vacuum
is much higher
and the tunneling into the true vacuum is much more suppressed.

\section*{Acknowledgement}
This work is partially supported by ``Priority Area: Supersymmetry and
Unified Theory of Elementary Particles (No.707)''(M.K. and T.Y.).
The work of T.W. is supported by Japan Society for the Promotion of Science.

\begin{thebibliography}{99}
\bibitem{SuperK} Y. Fukuda {\it et al.}
     [Superkamiokande Collaboration],
     \PLB{433}{9}{1998}; \PLB{436}{33}{1998}; \PRL{81}{1562}{1998}.
\bibitem{SeeSaw}T. Yanagida,
     {\it in} Proc. Workshop on the unified theory and
     the baryon number in the universe, (Tsukuba, 1979),
     {\it eds.} O. Sawada and S. Sugamoto,
     Report KEK-79-18 (1979);\\
     M. Gell-Mann, P. Ramond and R. Slansky,
     {\it in} ``Supergravity''
     (North-Holland, Amsterdam, 1979)
     {\it eds.} D.Z. Freedman and P. van Nieuwenhuizen.
\bibitem{Randall-Sundrum} L. Randall and R. Sundrum,
     \NPB{557}{79}{1999}.
\bibitem{Giudice-etal} G.F. Giudice, R. Rattazzi, M.A. Luty
     and H. Murayama,
     \JHEP{12}{027}{1998}.
\bibitem{Buchmuller}
     See, for a recent review,
     W. Buchm\"uller and M. Pl\"umacher,
     \PRT{320}{329}{1999};\\
     M. Pl\"umacher,
     \NPB{530}{207}{1998}.
\bibitem{Ellis}
     M. Yu. Khlopov and A.D. Linde,
     Phys. Lett. {\bf 138B}, 265 (1984);\\
     J. Ellis, G.B. Gelmini, J.L. Lopez, D.V. Nanopoulos and S. Sarker,
     Nucl. Phys. {\bf B373}, 399 (1992); \\
     M. Kawasaki and T. Moroi,
     Prog. Theor. Phys. {\bf 93}, 879 (1995).

\bibitem{Gherghetta}
     T. Gherghetta, G.F. Giudice and J.D. Wells,
     hep-ph/9904378;\\
     J.L. Feng and T. Moroi,
     hep-ph/9907319;\\
     see also Ref.\cite{Randall-Sundrum}.
\bibitem{Coleman} S.Coleman,
     \PRD{15}{2929}{1977};\\
     S. Coleman and F. De Luccia,
     \PRD{21}{3305}{1980}.
\bibitem{Linde}
     A.D. Linde,
     \NPB{216}{1983}{421}.
\bibitem{KolbTurner}
     See, for example, E.~Kolb and M.~Turner,{\it The Early Universe},
     (Addison-Wesley,1990)
\bibitem{Gleiser}
      M. Gleiser, E.W. Kolb and R. Watkins,
      \NPB{364}{411}{1991}.
\bibitem{Guth-Weinberg}
     A.H. Guth and E.J. Weinberg,
     \PRD{23}{876}{1981}.
\end{thebibliography}
\end{document}